\documentclass{aa}
\usepackage{txfonts}
\usepackage{graphicx}
\usepackage{natbib}
\bibpunct{(}{)}{;}{a}{}{,} 
\begin{document}
\newcommand{\chemin}{images_ELT2/}
\newcommand{\tf}{\widehat}
\title{Fundamental limitations on Earth-like planet detection with Extremely Large Telescopes}
\author{C. Cavarroc\inst{1}  \and A. Boccaletti\inst{1} \and P. Baudoz\inst{1} \and T. Fusco\inst{2} \and D. Rouan\inst{1}}
\institute{LESIA, Observatoire de Paris Meudon, 5 pl. J. Janssen, 92195 Meudon, France \\
\email{celine.cavarroc@obspm.fr} 
\and ONERA, BP 52, 29 avenue de la Division Leclerc, 92320 Ch\^atillon Cedex, France}
\offprints{C. Cavarroc}
\titlerunning{Fundamental limitations on Earth-like planet detection with ELTs}

\abstract{
We analyse the fundamental limitations for the detection of extraterrestrial planets with Extremely Large Telescopes. 
For this task, a coronagraphic device combined to a very high order wavefront correction system is required but not sufficient to achieve the $10^{-10}$ contrast level needed for detecting an Earth-like planet. The stellar residuals left uncorrected by the wavefront correction system need to be calibrated and subtracted. In this paper, we consider a general model including the dynamic phase aberrations downstream the wavefront correction system, the static phase aberrations of the instrument and some differential aberrations provided by the calibration unit. A rather optimistic case of a filled circular pupil and of a perfect coronagraph is elsewhere assumed.
As a result of the analytical study, the limitation mostly comes from
the static aberrations. Using numerical simulations we confirm this
result and evaluate the requirements in terms of phase aberrations to
detect Earth-like planets on Extremely Large Telescopes.

}

\maketitle
\section{Introduction}
Direct detection of Earth-like planets is a challenging issue, since
a contrast of $10^{-10}$ must be reached at close angular distance to a bright star.
The characterization of exoplanets constitutes the main objective of several projects which
are studied today, whether from the ground or from space (Darwin, TPF-I and TPF-C for instance). 
Among ground-based projects, Extremely Large Telescopes (ELTs) are very promising to 
significantly improve the angular resolution. To date, several concepts are considered like the Giant Magellan Telescope 
 \citep{johns-2004},  the Thirty Meters Telescope \citep{nelson-2005}, Euro50 \citep{euro50-2003} and the 100 m OWL telescope \citep{gilmozzi-2004}.

For detecting Earth-like planets, the use of a coronagraph combined to
an adaptive optics (hereafter AO) system is necessary but not sufficient. In presence of atmospheric turbulence we may not expect a contrast larger than $\sim10^{-6}$ even with a high-order AO (whatever the coronagraphic system is). Assuming an optimistic error budget for the AO, the Strehl ratio can be as large as 98\% (see section \ref{S-simu}). Thus, the coronagraph is feeded with an imperfect wavefront. Two percent of the stellar light is still escaping the coronagraph and is spread in a speckled halo. 

To enhance the contrast, a second step is mandatory to suppress this speckle noise. 
In this paper, we consider that the instrument is providing a good calibration and that the speckle attenuation is achieved with a simple subtraction of 2 images (one of which shows up the planet while the other does not). 
Due to the variability of the atmospheric phase distortion, the calibration must be done simultaneously. Here, for sake of generality we do not assume any particular system of calibration but several proposals were made in the past which take advantage of the planet properties like its spectral signature \citep{racine-1999,marois-2000}, its polarization \citep{baba-2003} or its coherence with respect to the star \citep{guyon-2004}. The differential spectral imaging was already implemented in TRIDENT \citep{marois-2005} and NACO-SDI \citep{close-2004}  and is also selected for the VLT Planet Finder \citep{beuzit-2004}.

Section \ref{S-formalism} presents the formalism based on a simple model. Numerical simulations (section \ref{S-simu}) were done to confirm the analytical results and to assess the sensitivity to the various phase aberrations. Important conclusions are drawn in section \ref{S-conclusion}.

\section{Formalism\label{S-formalism}}
\subsection{Assumptions\label{S-hyp}}
The following assumptions are very general and mostly optimistic with respect to the performances of the state-of-the-art instruments.

We consider a circular unobstructed pupil without segmentation. The
coronagraph is assumed perfect and is modeled by the subtraction of a
perfect pupil to the actual aberrated one (see Eq. \ref{E-ref} and \ref{E-A2}) in
order to remove totally the diffraction. It is achromatic, the star is perfectly centered and the Lyot stop is perfectly aligned.

Several coronagraphs which provide a total extinction of an on-axis unresolved point source with a zero aberration system were already proposed \citep{roddier-1997,rouan-2000,  baudoz-2000a, kuchner-2002}. But, to find the most suitable type of coronagraph it will be necessary to take into account the pupil defects of actual ELTs (segmentation, obstruction, cophasing errors between segments,...).

We consider the following contributions to the phase errors in the
wavefront: residual phase from AO correction ($\phi$) and static
aberrations ($\delta_C$) in the instrument upstream the coronagraph. This second type of
aberration is mandatory to be realistic, even if, actually, they are not totally
static but 'quasi static'  with a lifetime from a few seconds to a few minutes \citep{marois-2003}.

With these notations and assuming a perfect coronagraph, the complex amplitude $A_1$ in the pupil plane can be written as:
\begin{equation}\label{E-ref}
A_1=\Pi\left(\sqrt{E_C}-e^{\displaystyle {i\left(\phi+\delta_C\right)}} \right)
\end{equation}
with $\Pi$ the entrance pupil,
$E_C=e^{\displaystyle{-\sigma_{\phi+\delta_C}^2}}$ the instantaneous coherent energy and
$\sigma_{\phi+\delta_C}^2$ the spatial variance of the AO residual
phase and of the static aberration.

Assuming a high Strehl ratio (more than 95\%) i.e. considering that
wavefront errors are small (a few nanometers or tens of nanometers
RMS), a third order expansion can be made and is justified in the next
section. We can approximate $A_1$ with:
\begin{equation}\label{E-A1lin}
A_1=\Pi\left(-\frac{\sigma_{\phi+\delta_C}^2}{2}-i\left(\phi+\delta_C\right)+\frac{\left(\phi+\delta_C\right)^2}{2}+i\frac{\left(\phi+\delta_C\right)^3}{6}\right)
\end{equation}

\subsection{Instantaneous residual intensity}
We consider here two images taken simultaneously using two channels
downstream the coronagraph to calibrate the residual speckle pattern. 
A potential high source of mismatch in between the
two images is the non common path wavefront aberration since the light goes through two differential optical paths. 
So, we will consider a 'reference' channel having only $\phi$ and $\delta_C$, the amplitude of which
is described by Eq. \ref{E-ref} and a second channel which, in
addition to $\phi$ and $\delta_C$, includes a non common path
aberration $\delta_{NC}$ with an amplitude defect of the same order as $\phi$ and $\delta_C$. Its amplitude in the pupil plane
can be written by:
\begin{equation}\label{E-A2}
A_2=\Pi\left(\sqrt{E_C}-e^{\displaystyle{i\left(\phi+\delta_C\right)}}\right)e^{\displaystyle{i\delta_{NC}}}
\end{equation}

As for Eq. \ref{E-ref} a third order expansion gives:
\begin{equation}\label{E-A2lin}
A_2=A_1+\Pi\left(-i\frac{\sigma_{\phi+\delta_C}^2}{2}\delta_{NC}+\left(\phi+\delta_C\right)\delta_{NC}+i\frac{\left(\phi+\delta_C\right)^2\delta_{NC}}{2}\right)
\end{equation}

The intensity $I_2$ in the detector plane corresponds to the square
modulus of the Fourier transform of $A_2$ (neglecting terms beyond
third order). We will denote Fourier transforms by a hat
accent. The star $\star$ represents the convolution operator.

\begin{eqnarray}
I_2&=&\mid\tf{A_2}\mid^2\\
&=&I_1+2\Im\left[\left(\tf{\Pi}\star\left(\tf{\phi}+\tf{\delta_C}\right)\right).\left(\tf{\Pi}\star\left(\tf{\phi}+\tf{\delta_C}\right)\star\tf{\delta_{NC}}\right)^*\right]\nonumber
\end{eqnarray}
with $^*$ the complex conjugation and $\Im[\ ]$ the imaginary
part of the complex quantity. $I_1$ is the intensity
of the reference image ($I_1=\mid\tf{A_1}\mid^2$).

Subtracting $I_1$ from $I_2$, we obtain:
\begin{eqnarray}\label{E-diff}
I_{NC}=I_2-I_1&=&
2\Im\left[\left(\tf{\Pi}\star\tf{\phi}\right).
\left(  \tf{\Pi}\star\tf{\phi}\star\tf{\delta_{NC}}\right)^* \right]\nonumber\\
&+&2\Im\left[\left(\tf{\Pi}\star\tf{\phi}\right).
  \left(\tf{\Pi}\star\tf{\delta_C}\star\tf{\delta_{NC}}\right)^*\right] \\
&+&2\Im\left[\left(\tf{\Pi}\star\tf{\delta_C}\right).
  \left(\tf{\Pi}\star\tf{\phi}\star\tf{\delta_{NC}}\right)^* \right]\nonumber\\
&+&2\Im\left[\left(\tf{\Pi}\star\tf{\delta_C}\right). \left(\tf{\Pi}\star\tf{\delta_C}\star\tf{\delta_{NC}}\right)^*\right]\nonumber
\end{eqnarray}
$I_{NC}$ refers to the instantaneous residual intensity after subtraction.

\begin{figure}
  \resizebox{\hsize}{!}{\includegraphics{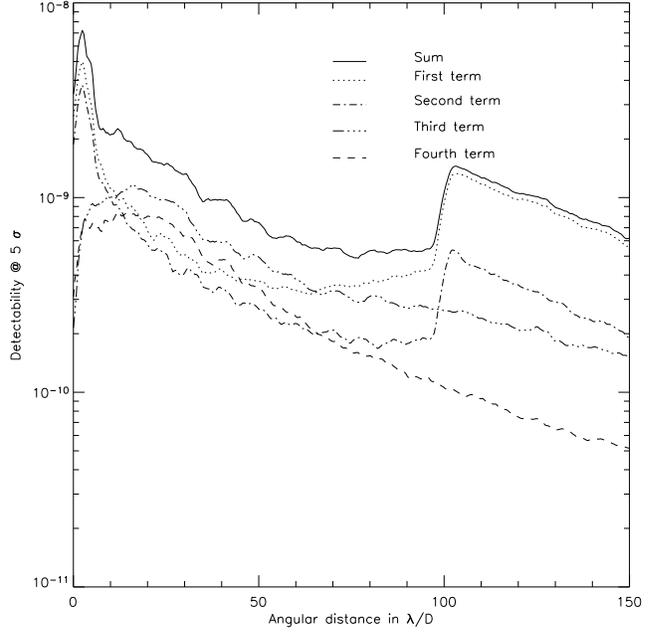}}
 \caption{Radial profile of the standard deviation (azimuthally averaged) of each term in equation
   \ref{E-diff} for an instantaneous image (1 speckle pattern). Simulation
   parameters are given in section \ref{S-simu}.}
 \label{I-0}
\end{figure}
$\Pi$, $\phi$, $\delta_C$ and $\delta_{NC}$ are real, so their Fourier
transforms are hermitian: their real part is even and their imaginary
part is odd. Thus, the resulting image $I_{NC}$, which is made only
with imaginary parts, is antisymmetric (providing the expansion is stopped at the third order).

Equation \ref{E-diff} has only terms of the third order. 
It is therefore obvious that the third order expansion in Eq. \ref{E-A1lin} and \ref{E-A2lin} are required to obtain $I_{NC}\neq 0$.
All terms which classically dominate imaging are removed in the 2-channel subtraction process. Indeed, an image made with a perfect coronagraph has neither residual constant nor first order terms, as
developed in \citet{boccaletti-2002} and \citet{bloemhof-2004} in the case of the
Four Quadrant Phase Mask. Moreover, the interest of coronagraphic image subtraction
lies in removing terms of degree two which are studied in
\citet{perrin-2003} and \citet{siva-2002} in a direct imaging case.

The terms of equation \ref{E-diff} varie linearly with
$\tf{\delta_{NC}}$. Because of the linearity of the Fourier
transform, it means that a linear variation in $\delta_{NC}$ induces a
linear variation of the level of the residual intensity.
Each term of Eq.  \ref{E-diff} are shown in Fig. \ref{I-0} for a single instantaneous image.

The approximation (third order expansion) of the  Eq. \ref{E-A1lin} and \ref{E-A2lin} is valid when the aberrations $\phi$, $\delta_C$ and $\delta_{NC}$ have similar orders of magnitudes. In actual cases, 
$\delta_{NC}$ will differs by 1 or 2 orders of magnitude which is changing the order of the expansion but 
Eq. \ref{E-diff} will remain unchanged (as confirmed in Fig. \ref{I-4}).

\section{Numerical simulations\label{S-simu}}
We carried out numerical simulations to illustrate and confirm the analytical study.
Simulations were performed for a thirty
meters telescope, using arrays of 2048$\times$2048 pixels with a sampling of 3.6 cm
by pixel at the pupil plane and for a wavelength $\lambda=1.63\ \mathrm{\mu m}$. We considered a filled circular aperture.

The atmospheric phase screens are generated with a tool based on the approach first introduced by F. Rigaut \citep{Rigaut-p-98,Jolissaint-p-01}.  Analytical  expressions of the spatial power spectral density (PSD) of the residual phase  are obtained for various  errors affecting an AO system (fitting, aliasing, temporal, noise, anisoplanatism ...). The resulting global AO PSD (sum of the individual PSD of each error source) is used to compute AO corrected phase screens (independent realizations of the global AO PSD). Recent works \citep{Rconan-p-03} allowed to include a correct model of  AO closed loop temporal and noise effects (assuming an intregator law) as well as the differential refraction effects. 

In the following, we consider a Shack-Hartmann wavefront sensor with 200 actuators across the telescope diameter. The seeing is 0.75 arcsecond at $\lambda=0.5\mu$m and the outer scale is 20 m. 
We assumed a 5th magnitude star and a loop frequency of 2.5 kHz. The AO PSD includes fitting errors, servo lag and the photon noise on the wavefront sensor. These AO parameters provide a very good wavefront correction and a Strehl ratio of 98\% which is definitely a great challenge for ELTs.

The static phase aberrations $\delta_C$ and $\delta_{NC}$ are described with a PSD which varies as $f^{-2}$ (with $f$ the spatial frequency) as it is usually the case for standard optical components \citep{duparre-2002}. In the case of high contrast imaging with a coronagraph, special care is needed to reduce the impact of low spatial frequencies. We therefore assumed that the PSD at low frequencies was already improved and that it is flat in the range $0<f<f_C/4$, with $f_C$ the cut-off frequency of the AO defined by the number of actuators across the pupil.
As a first guess we adopt an amplitude of 20 nm RMS and 0.1 nm RMS for  $\delta_C$ and $\delta_{NC}$ respectively.
The impact of the amplitude and the PSD of the phase aberrations are further investigated in sections \ref{S-value} and \ref{S-prof}.

\subsection{ Short exposure image}
\begin{figure}
 \resizebox{\hsize}{!}{\includegraphics{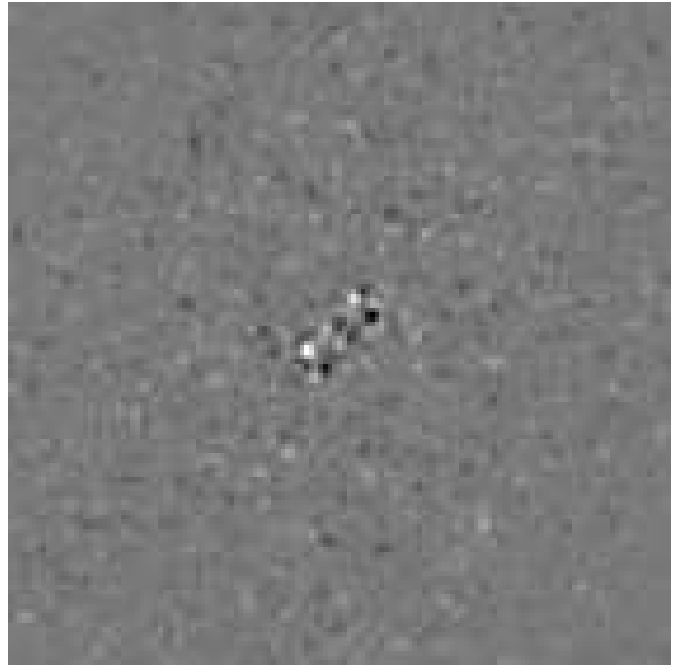}}
 \caption{Central part of an instantaneous image (fov=$54\lambda/D$). 
 The antisymmetrical speckle pattern is evidenced. The
   simulation was done for $\delta_C=20$ nm RMS and $\delta_{NC}= 0.1$
   nm RMS and with a power square law spectrum truncated at $f_c/4$ ($f_c$ is 
   the cut-off frequency of the AO system).}
 \label{I-1}
\end{figure}

Let's study first a single short exposure coronagraphic image subtracted with a calibration image. Its theoretical expression is given by
Eq. \ref{E-diff}. Figure \ref{I-1} shows the central part of
an instantaneous simulated image after subtraction ($I_{NC}=I_2-I_1$). Positive
speckles are white while negative ones are black. The intensity distribution in the focal plane is clearly anti centrosymmetrical as expected from Eq. \ref{E-diff}.

Unlike direct imaging, speckle pinning \citep{bloemhof-2001} which appears when a phase defect is multiplied by the Fourier transform of the pupil ($\tf{\Pi}$) is not occurring here (see Eq. \ref{E-diff}) thanks to the coronagraph effect and the image subtraction.
Thus, residual speckles are randomly distributed in the image and not only pinned about the rings of the point spread function.

\begin{figure}
  \resizebox{\hsize}{!}{\includegraphics{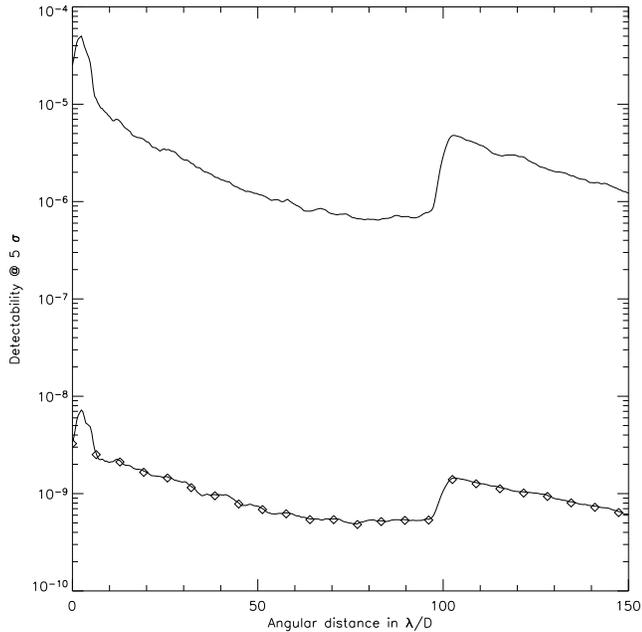}}
 \caption{Comparison between analytical (solid lines) and simulated
   curves (diamonds) calculated to achieve a 5$\sigma$ detection. The
   upper curve corresponds to the raw coronagraphic image $I_1$.  The
   bottom curve corresponds to standard deviation of the instantaneous image after subtraction, $I_{NC}$ (Eq. \ref{E-diff}). The static aberrations have the same amplitude and PSD as in Fig.  \ref{I-1}. }
 \label{I-2}
\end{figure}

 Figure \ref{I-2} compares radial profiles of the image before and after
 subtraction for a single exposure. The agreement between simulated and analytical profiles is perfect, indicating that the approximation we did in section \ref{S-formalism} is valid at this level of contrast.

\subsection{Convergence over time\label{S-cv2}}
\begin{figure}
  \resizebox{\hsize}{!}{\includegraphics{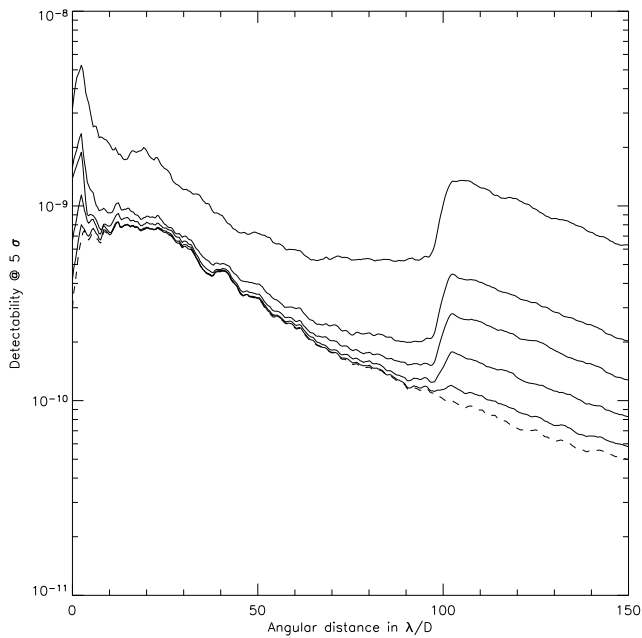}}
  \caption{Detectability at 5$\sigma$ as a function of the number of realizations N.
 The different solid lines represent the azimuthal standard deviation of the quantity $<I_{NC}>$ and are calculated with 9, 27, 90 and 460 decorrelated  speckle patterns (from top to bottom respectively). The theoretical limit is derived from Eq. \ref{E-limit} (dashed line). The static aberrations have the same amplitude and PSD as in Fig.  \ref{I-1}. }
  \label{I-3}
\end{figure}

Long exposure image is generated by co-adding N decorrelated
instantaneous images while static aberrations remain unchanged.
The final image delivered by the simulation is $<I_{NC}>$ where $<>$ denotes temporal averaging over the N realizations. In the following, we analyze this simulation result in terms of detectability of a point source. For that, we calculated the standard deviation in the image $<I_{NC}>$ as explained in \citet{boccaletti-2004}. This operation is noted $\sigma_{\theta}$ and is measured on a sample of pixels at equal distance to the star.

In Fig. \ref{I-3}, the radial profiles are calculated for increasing values of N from a single exposure to N=460.
This simulation shows that the detectability 
converges towards the standard deviation of the constant term of Eq. \ref{E-diff} (dashed line in Fig. \ref{I-3}):
\begin{equation}\label{E-limit}
\sigma_{\theta}\left(<I_{NC}>\right) = \sigma_{\theta}\left(2\Im\left[\left(\tf{\Pi}\star\tf{\delta_C}\right). \left(\tf{\Pi}\star\tf{\delta_C}\star\tf{\delta_{NC}}\right)^*\right]\right)
\end{equation}

The other terms containing $\phi$ average to an azimuthally constant pattern over time when a large
number of phase screens are simulated (N$\gtrsim 500$). Therefore, the
detectability only depends on the static aberrations upstream
($\delta_C$) and downstream ($\delta_{NC}$) the coronagraph and
Eq. \ref{E-limit} gives the fundamental limit of contrast with the
model we have considered. The detectability varies linearly with $\delta_{NC}$ but quadratically with $\delta_C$.
In addition, it is clear that a simulation where the static aberrations are omitted leads to a complete averaging of the speckle pattern over time which is actually unrealistic.

In the following we will only consider Eq. \ref{E-limit} as the fundamental limitation to assess the influence of $\delta_C$ and $\delta_{NC}$.

\subsection{Influence of static aberration amplitudes\label{S-value}}
\begin{figure}
  \resizebox{\hsize}{!}{\includegraphics{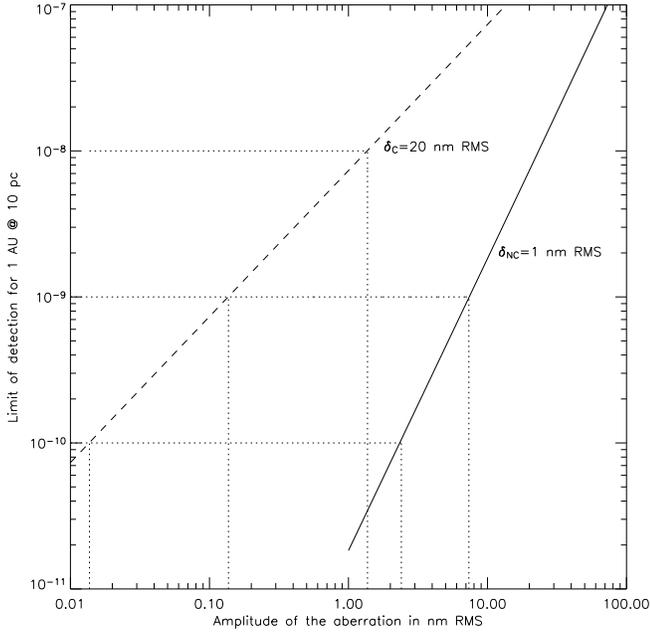}}
  \caption{Detectability at $5\sigma$ for an angular separation  
 of $9 \lambda/D$ corresponding to 1 AU at 10 pc.
The solid line stands for the impact of  $\delta_C$ when $\delta_{NC}$ is set to 1 nm RMS and conversely the dashed line represents the impact of $\delta_{NC}$ when $\delta_C$ is set to 20 nm RMS. The PSD of aberrations is identical to Fig. \ref{I-1}.}
  \label{I-4}
\end{figure}

The influence of the amplitude of the static aberrations is shown in Fig. \ref{I-4} for a radial distance in the field corresponding to a 1 AU orbit located at 10 pc.  
The impact of common path and non common path static aberrations is clearly different.  
The dependence of the detectability varies quadratically with $\delta_C$ and linearly with $\delta_{NC}$ as expected from Eq. \ref{E-limit}. The influence of common static aberrations upstream the coronagraph is therefore more important.

To reach a contrast of $10^{-10}$ at 0.1" as required to detect an Earth-like planet located at 10 pc, the non common static aberrations downstream the coronagraph must be controlled to 0.01 nm RMS if common aberrations are set to 20 nm RMS. Reversely, one can tolerate non common aberrations of 1 nm RMS if the common aberrations upstream the coronagraph are lowered down to 3 nm RMS. Common aberrations are dominant and must be reduced as far as possible.

\begin{figure}
  \resizebox{\hsize}{!}{\includegraphics{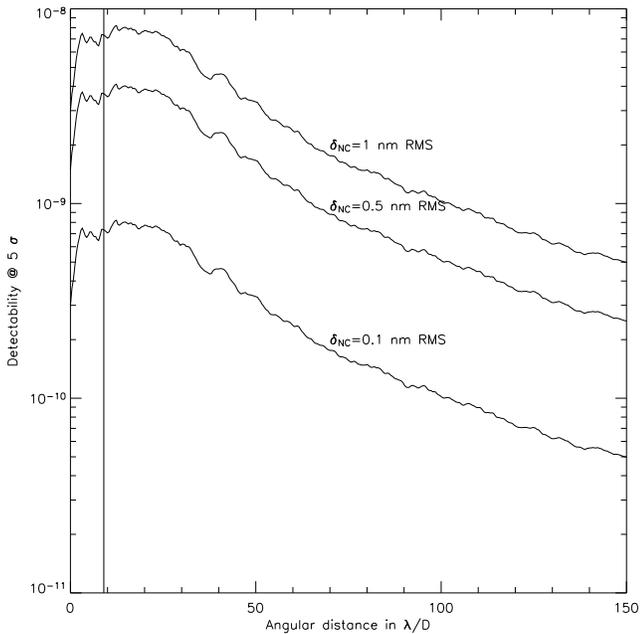}}
  \caption{Profiles of detectability at 5 $\sigma$ with a common
    path aberration $\delta_C$ fixed to 20 nm RMS for several
    values of $\delta_{NC}$. The vertical line shows the position of a 1 AU orbit at 10 pc.}
  \label{I-5}
\end{figure}

Figure \ref{I-5} shows the radial profile of the detectability at 5 $\sigma$ for different amplitudes of $\delta_{NC}$ assuming
$\delta_C$ is set to 20 nm RMS (a challenging but realistic value). The linearity is again evidenced here.

\subsection{Influence of static aberration PSD\label{S-prof}}

\begin{figure}
  \resizebox{\hsize}{!}{\includegraphics{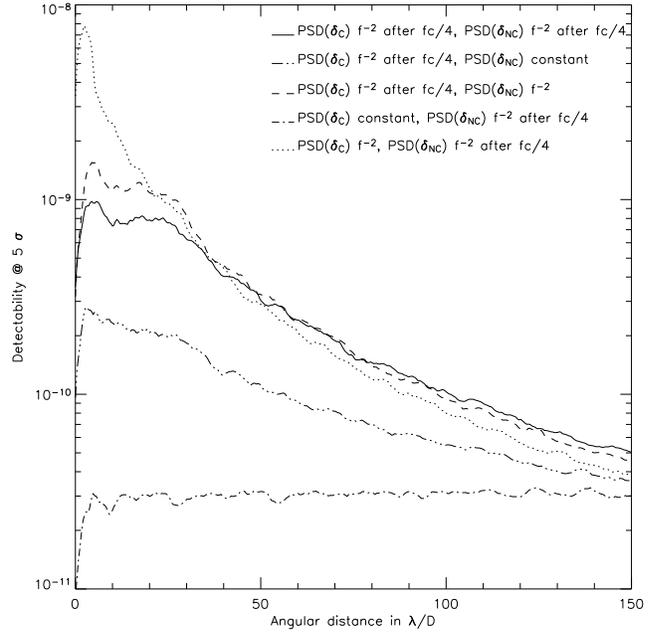}}
  \caption{Radial profile at 5 $\sigma$ for several combinations of PSD. For each case, the amplitude of 
  $\delta_C$ and $\delta_{NC}$ are set to 20 and 1 nm RMS. The solid line stands for the standard condition we have considered in previous sections to evaluate the detectability.}
  \label{I-6}
\end{figure}

Assuming the amplitudes of $\delta_C$ and $\delta_{NC}$ are set to 20 and 0.1 nm RMS respectively, we 
analyzed the influence of the Power Spectrum Density in Fig. \ref{I-6}.
The solid line corresponds to our standard conditions, where the PSD of $\delta_C$ and $\delta_{NC}$ is flat between 0 and $f_C/4$ and then varies as $f^{-2}$ for $f>f_C/4$. The degradation is not dramatic if the PSD of $\delta_{NC}$ varies like $f^{-2}$ at all frequencies (dashed line) which confirms the low sensitivity to this parameter. However, when the PSD of $\delta_C$ is not flat at low frequencies (dotted line) the detectability decreases by a factor 10 at small angular distances ($\sim 3\lambda/D$) and meets the solid line at $30\lambda/D$. Better results are obtainable if the PSD of static aberrations can be made flat at all frequencies (dot-dashed and dot-dot-dot-dashed lines). A significant improvement of the detectability is achievable when for instance the PSD of $\delta_C$ is flat even if that of  $\delta_{NC}$ is quadratic (dot-dashed line).

Again, the analysis of the PSD shows that $\delta_C$ is the dominant source of degradation and that the low frequencies must be appropriately controlled to achieve a good starlight reduction.

\subsection{Photon noise and influence of the diameter}\label{S-photon}
Another fundamental limitation we finally have to take into account is the
photon noise. It is important to know if the contrast level in Eq. \ref{E-limit} is achievable in a reasonable amount of time. We assumed a 5th magnitude G2V star at 10pc
observed with a 30 m telescope, a spectral resolution of $\lambda/\Delta\lambda=5$ and an overall transmission of 5\%.

\begin{figure*}
\centerline{
  \includegraphics[width=8.5cm]{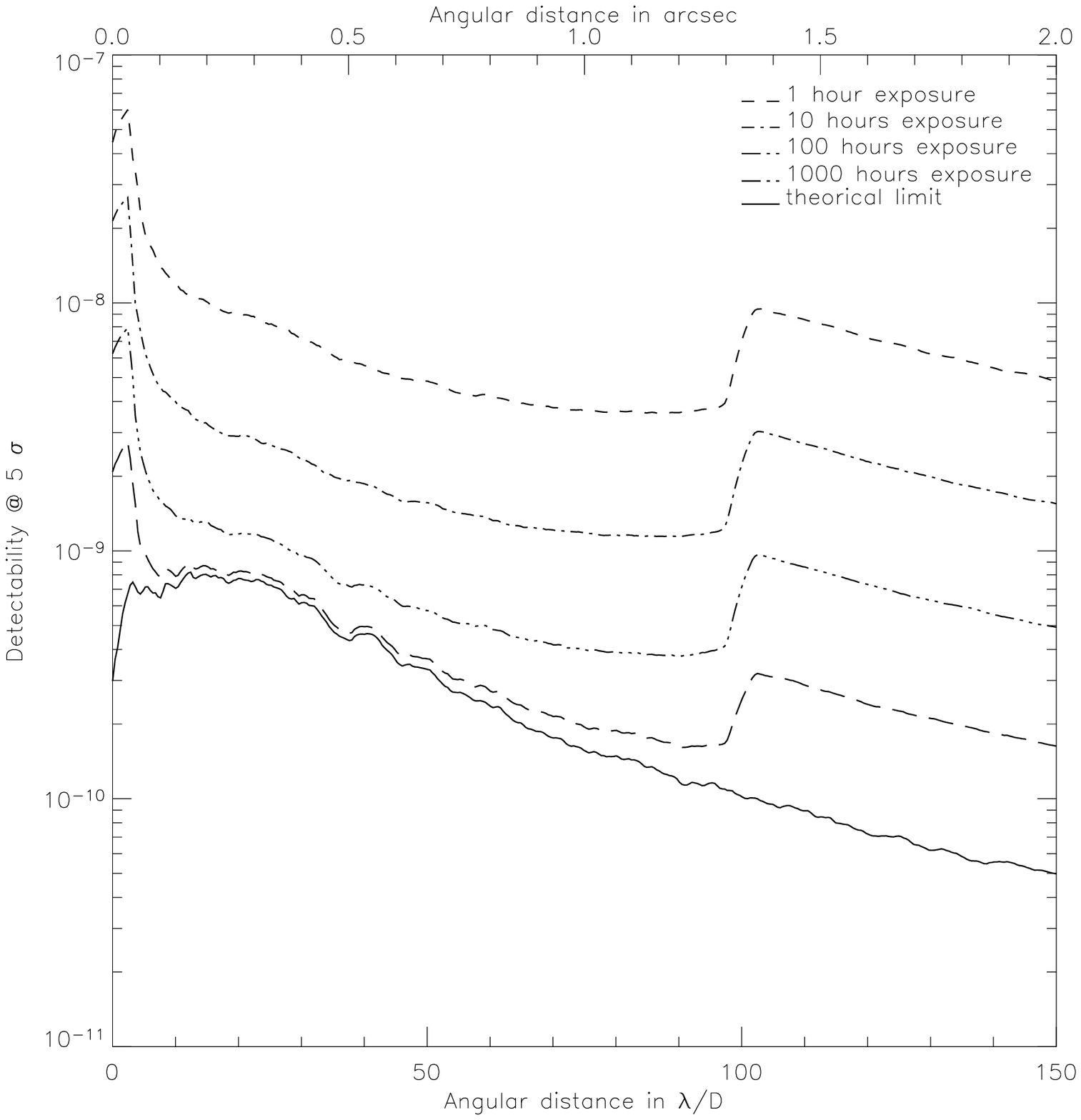}
  \includegraphics[width=8.5cm]{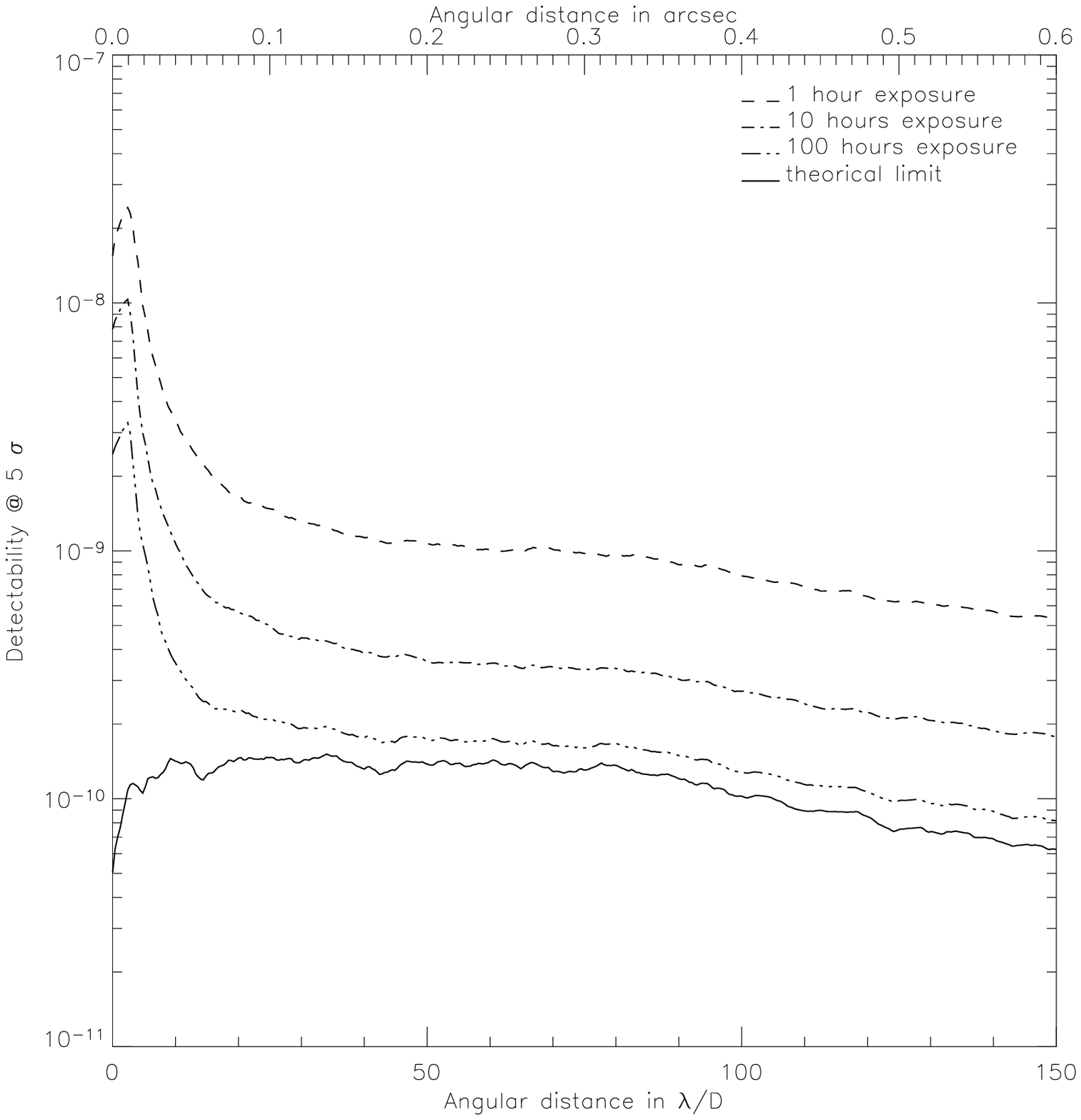}} 
    \caption{Radial profile of the detectability at 5 $\sigma$ for
a 30 m telescope (left) and a 100 m telescope (right). The lower curve corresponds to the theoretical
    limit ($\delta_C=$20 nm RMS and $\delta_{NC}=$0.1 nm RMS) for  an infinitely long exposure. Static aberrations scale with the diameter.}
  \label{I-7}
\end{figure*}

Figure \ref{I-7} (left) gives the limit of detectability for several exposure
times from 1 to 1000 hours and compares it to the expression of
Eq. \ref{E-limit} (solid line). The photon noise is included in the
image $I_1$ and $I_2$ and therefore the detectability has a radial profile impressed by the distribution of the atmospheric turbulence residual $\phi$. As the integration time increases, the detectability level and its profile converge to the expression given in Eq. \ref{E-limit}. 
A few hundreds of hours are required to achieve a 5$\sigma$ contrast of $8.10^{-10}$ at a separation of 0.1".

A more favorable case is achieved with a 100 m telescope (Fig. \ref{I-7} right) assuming static aberrations have the same amplitudes ($\delta_C=$20 nm RMS and $\delta_{NC}=$0.1 nm RMS) and a cut-off frequency of $f_c/4$. The AO system has the same actuators density than for a 30 m so the linear number of actuators amounts to 667. In that situation, the influence of static aberrations is greatly relaxed  and a contrast of $2.10^{-10}$ is achievable in about 100 hours at 0.1". The PSDs of static aberrations have the same integral for the 30 m and the 100 m telescopes but the cut-off frequency is 3 times larger (83$\lambda/D$ instead of 25$\lambda/D$) so that the amplitude of low frequencies is reduced and finally, the limit of detection is improved by a factor of 5.

\begin{figure}
  \resizebox{\hsize}{!}{\includegraphics{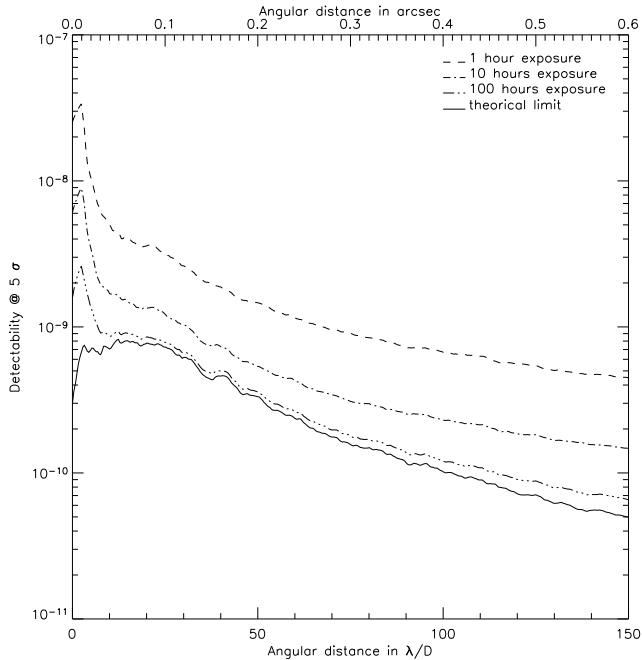}} 
    \caption{Radial profile of the detectability at 5 $\sigma$ for a 100 m telescope. 
    Static aberrations are included in the instrument rather than on the telescope and are independent of the diameter. The lower curve corresponds to the theoretical limit ($\delta_C=$20 nm RMS and $\delta_{NC}=$0.1 nm RMS) for  an infinitely long exposure.}
  \label{I-8}
\end{figure}

However, it is more realistic to consider that the static aberrations are located inside the instrument the size of which does not depend on the telescope diameter. In that case the cut-off frequency is identical for the 30 m and the 100 m telescopes so as the integral of the PSD and the limit of detection is found to be $8.10^{-10}$ as in Fig. \ref{I-7} (left). Nevertheless, the integration time to achieve this contrast is about 100 hours because the natural contrast of the PSF at 0.1" is improved with respect to the 30 m telescope (Fig. \ref{I-8}). 


\section{Comparison with previous studies}\label{S-discussion}
Several analytical studies were carried out in the past few years to assess the detectability of Earth-like planets with ELTs \citep{angel-2003, lardiere-2004, chelli-2005}. As shown in the previous section, the integration time depends on several parameters and in particular on the PSF contrast which quantifies the level of the residual light at a given angular separation on a short exposure. 
If the speckle noise is neglected and static aberrations are omitted as in \citet{lardiere-2004}, the signal to noise ratio is found to be only limited by the photon noise in the PSF halo. In that condition, the integration time varies linearly with the PSF contrast $C(\theta)$ where $\theta$ is the angular separation of interest. The critical parameter $C(\theta)$ is derived from the PSD of the residual phase left uncorrected by the AO system and therefore varies according to the different authors depending on the amplitude of the errors which are considered (servo lag, photon noise on the WFS, fitting errors, ...).

In \citet{lardiere-2004}, this PSF contrast amounts to $5.10^{-8}$ for a 100 m telescope which allows the detection of a $10^{-10}$ contrast in 3.5 hours at 5$\sigma$ assuming a stellar flux of $10^{10}ph/s$. Similar performances were obtained by \citet{chelli-2005} (where the speckle noise is treated correctly) and \citet{angel-2003} which both adopted the same value of $C(\theta)$.

In our study, we found significantly larger integration times even with a 100 m telescope for which the static aberrations are set to a very low level (Fig. \ref{I-7} right). 
This discrepancy precisely originates from the performance of the AO system since our phase screen simulation yields a PSF contrast at 0.1" of $1.5.10^{-6}$ for a 30 m telescope and $5.10^{-7}$ for a 100 m which is corresponding in the Lardiere et al model to integration times of 100 hours and 35 hours respectively. The numerical simulation including photon noise (section \ref{S-photon}) and a more realistic model of the phase aberrations is even less favorable and predicts an integration time of about 100 hours on 100 m telescope to achieve a contrast of $2.10^{-10}$ at 5$\sigma$.

From this rough comparison, it is clear from that beyond the problem of static aberrations,
the potential of ELTs to detect Earth-like planets mostly resides in the performance of the AO system.

\section{Conclusion}\label{S-conclusion}
In this study, we considered an actual system, taking into account
some defects due to the AO residuals $\phi$, the common path aberrations
$\delta_C$ between the AO system and the coronagraph and the non common path
aberrations $\delta_{NC}$ downstream the coronagraph. 
From that simple but realistic model we draw important conclusions: 

\begin{itemize}
\item{An actual system has $\phi>\delta_C>\delta_{NC}\neq0$. Any study where $\delta_C$ is neglected conducts to unrealistic results where the variance of the residual intensity necessarily converge to 0 on infinitely long exposure.}
\item{The fundamental limit for an infinitely long exposure is constant and is given by: \\
$\sigma_{\theta}\left(<I_{NC}>\right) = \sigma_{\theta}\left(2\Im\left[\left(\tf{\Pi}\star\tf{\delta_C}\right). \left(\tf{\Pi}\star\tf{\delta_C}\star\tf{\delta_{NC}}\right)^*\right]\right)$}
\item{This detectability has a quadratic dependence in $\delta_C$ and a linear dependence in $\delta_{NC}$ but does not depend on the AO performance because $\phi$ averages to 0 over time.}
\item{Improvement of the detectability can be obtained if the power spectrum density of the static phase aberrations is decreased, especially at low and mid frequencies.}
\item{To achieve a 5$\sigma$ contrast of $10^{-9}$ with a 30 m telescope when the common aberrations are 20 nm RMS, the calibration must be performed at a level of $\delta_{NC}=$0.1 nm RMS.  To achieve a contrast of $10^{-10}$ this level becomes extremely challenging: 0.01nm RMS.}
\item{An increasing of the telescope size doesn't change significantly the result and the same level of limitation holds for a 30 m and a 100 m telescope as long as the static aberrations originate from the instrument rather than the primary mirror. However, a larger telescope will result in a smaller integration time.}
\item{Beyond the limitation from static aberrations, the signal to noise ratio depends on the level of the PSF halo. Therefore, the performance of AO systems for ELTs is a critical point which must be addressed thoroughly. }
\item{Contrary to previous studies, it is shown that the feasibility of Earth-like planet detection with ELTs is questionable, for the reasons mentioned herein.}
\end{itemize}

It is worth to remind here that the present study did not take into account the telescope pupil segmentation and obscuration, the pointing of the source onto the coronagraph, the alignment of the pupil onto the Lyot stop and the defects of the coronagraph (chromatism for instance). Background noise and Flat Field will be also critical. An actual system must consider all these sources of error to derive realistic performance as it was done for the case of MIRI on JWST \citep{baudoz-2005} and VLT Planet Finder \citep{beuzit-2004}. Scintillation is another source of error to consider in the case of large ground-based telescopes when dealing with a $10^{-10}$ level. 

As a final conclusion, it is obvious in regard to the present study that the detection (and even more the characterization) of Earth-like planets from the ground with Extremely Large Telescopes will be extremely challenging. A thorough analysis of the system as a whole (telescope + AO system + instrument) is strongly recommended to tackle the many and probably unknown sources of error.

\bibliographystyle{aa}
\bibliography{liste3}

\end{document}